# Precise *dd* excitations and commensurate intersite Coulomb interactions in the dissimilar cuprate $YBa_2Cu_3O_{7-y}$ and $La_{2-x}Sr_xCuO_4$

Shih-Wen Huang,[1,2,3,4*] L. Andrew Wray[5*], Yu-Cheng Shao,[6,3] Cheng-Yau Wu,[7] Shun-Hung Wang,[7] Jenn-Min Lee,[2] Y-J. Chen,[3] R. W. Schoenlein,[8,4] C. Y. Mou,[9,10,11,12], Yi-De Chuang[3] and J.-Y. Lin,[7,13*]

[1] *Swiss Light Source, Paul Scherrer Institut, CH5232 Villigen PSI, Switzerland*

[2] *MAX IV Laboratory, Lund University, P. O. Box 118, 221 00 Lund, Sweden*

[3] *Advanced Light Source, Lawrence Berkeley National Laboratory, Berkeley, CA 94720, USA*

[4] *Materials Sciences Division, Lawrence Berkeley National Laboratory, Berkeley, CA 94720, USA*

[5] *Department of Physics, New York University, New York, NY 10003, USA*

[6] *National Synchrotron Radiation Researc Center, Hsinchu 30076, Taiwan*

[7] *Institute of Physics, National Yang Ming Chiao Tung University, Hsinchu 30010, Taiwan*

[8] *SLAC National Accelerator Laboratory, Menlo Park, California 94025, USA*

[9] *Physics Division, National Center for Theoretical Sciences, P.O. Box 2-131, Hsinchu, Taiwan.*

[10] *Center for Quantum Technology and Department of Physics, National Tsing Hua University, Hsinchu, 300, Taiwan*

[11] *Institute of Physics, Academia Sinica, Taipei 11529, Taiwan*

[12] *Department of Physics, National Tsing Hua University, Hsinchu 30043, Taiwan*

[13] *Center for Emergent Functional Matter Science, National Yang Ming Chiao Tung University, Hsinchu 30010, Taiwan*

**Abstract:**

Using high-resolution extreme ultraviolet resonant inelastic X-ray scattering (EUVRIXS) spectroscopy at Cu *M*-edge, we observed the doping dependent spectral shifts of inter-orbital (*dd*) excitations of $YBa_2Cu_3O_{7-y}$ and $La_{2-x}Sr_xCuO_4$. With increasing hole doping level from undoped to optimally doped superconducting compositions, the leading edge of *dd* excitations is found to shift towards lower energy loss in a roughly linear trend that is irrespective to the cuprate species. The magnitude of energy shift can be explained by including a 0.15 eV Coulomb attraction between Cu $3d(x^2-y^2)$ electrons and the doped holes on the surrounding oxygens in the atomic multiplet calculations. The consistent energy shift between distinct cuprate families suggests that this inter-site Coulomb interaction energy scale is relatively material-independent, and provides an important reference point for understanding charge density wave phenomena in the cuprate phase diagram.

**I. Introduction**

Even after more than three decades of intense research on cuprates, the mechanism of their high temperature superconductivity (HTSC) remains heavily debated. Generic features such as the antiferromagnetic (AF) Mott insulating state in the parent compounds, the ubiquitous charge and spin orders/stripes, the pseudogap, the formation of Zhang-Rice singlet (ZRS) upon hole doping, the salient Fermi surface topology, and the non-Fermi liquid behavior, manifest a universal phase diagram that has been seen several refinements [1-3]. However, contrasting properties among different cuprate families complicate the identification of HTSC mechanism [4, 5]. Taking La$_{2-x}$Sr$_x$CuO$_4$ (LSCO) and YBa$_2$Cu$_3$O$_{7-y}$ (YBCO) as an example. To name a few, LSCO has a simpler CuO$_2$ plane structure formed by corner-shared CuO$_6$ octahedra, whereas YBCO has CuO$_5$ square pyramids and additional chain layers that serve as the charge reservoir (Fig. 1(a)). LSCO generally has a lower maximum transition temperature ($T_{c,\,\text{max}}$) compared to YBCO and other non-214 systems [6]. Their charge order wave vectors also display opposite doping dependence that regards/disregards the commensurate $q = 0.25$ value [7-10]. In light of these vastly different properties, it is important to examine the energetic parameters used in theoretical models for explaining HTSC. It has been suggested that the essential ingredients of HTSC mechanism may be drawn from the energy scales associated with inter-orbital ($dd$) excitations, hence an accurate determination of these energy scales will be beneficial for validating relevant theoretical proposals [11-14].

Another elusive parameter in the multi-band Hubbard model is the inter-site Coulomb interaction $U_{pd}$, which is the Coulomb attraction between Cu $3d(x^2-y^2)$ electron and the ZRS on the neighboring copper-oxygen plaquettes [15, 16]. This $U_{pd}$ parameter plays a key role in the dynamics that follows the charge transfer excitation, making it indispensable for describing certain optical and resonant inelastic X-ray scattering (RIXS) excitations [17-21]. It is also a significant reference point for understanding the electron-hole Coulomb interaction that is at the root of charge density wave phenomena. $U_{pd}$ also contributes to the effective interaction between ZRS and doublons, where the latter one plays an increasingly important role in strongly correlated electron systems and their transient dynamics [22,23]. Interestingly, there has never been a convincing experimental determination of $U_{pd}$.

Novel spectroscopies have been used to explore the essential phenomenology of cuprates and elucidate these exotic and still puzzling features, and RIXS is one of these promising techniques [24-26]. RIXS is element-specific and can be used to directly probe elementary excitations like magnons, charge transfer (CT), and inter-orbital (*dd* and *ff*) excitations [27]. Especially, RIXS employed in the extreme ultraviolet (EUV) energy regime (*M*-edge resonances of 3*d* transition metals) can have a much higher energy resolution than that in soft X-ray regime (*L*-edge resonances of 3*d* transition metals). It also yields cleaner spectral profiles for *dd* excitations due to the suppressed shake-up and charge-transfer excitations, thus is an ideal tool to address the aforementioned questions.

## II. Experiments

We performed EUV-RIXS measurements on LSCO and YBCO films around Cu *M*-edge resonances at beamline 4.0.3 (MERLIN) MERIXS endstation at the Advanced Light Source (ALS), Lawrence Berkeley National Laboratory. All spectra were recorded at room temperature using an X-ray spectrometer placed at 90º scattering angle relative to the incident X-ray beam [28]. The intensity of scattered X-rays recorded by a CCD detector on the spectrometer was normalized to ensure that the intensity in final spectra accurately represents the density of scattered photons per unit energy [29]. During the measurements, the sample was maintained at

20º incidence angle relative to its surface normal with the Cu-O bond direction in the horizontal scattering plane. The photon polarization was kept in this plane (π polarization) to suppress the strong elastic peak, see Fig. S1 in Supplementary Materials [30]. The full width at half maximum (FWHM) of elastic peak was 21 meV (Fig. 2(a)). To reveal the subtle shift in the leading edge of *dd* excitations in RIXS spectra, a precise determination of their energies with respect to the strong elastic peak is crucial. We looked at the elastic peak position on the detector as we varied the incident photon energy and used a 3$^{rd}$ order polynomial fitting to obtain a conversion between detector pixel number and photon energy, thereby converting the emission energy of RIXS spectra to the energy loss. This approach allowed us to determine the energy loss scale better than 2 meV, roughly 10% of the energy resolution determined from the FWHM of elastic peak. Exact steps for this conversion are listed in Supplementary Materials [30].

The (001) oriented high quality YBa$_2$Cu$_3$O$_{7-y}$ (YBCO) and La$_{2-x}$Sr$_x$CuO$_4$ (LSCO) thin films were grown on (100) SrTiO$_3$ and LaSrAlO$_4$ substrates by pulsed laser deposition (PLD) method, respectively. A KrF excimer laser (λ = 248 nm) with 5 Hz repetition rate and 5 J/cm$^2$ energy density was used to ablate the polycrystalline targets. For YBCO (LSCO), the growth was performed at 740 – 770 K (770 – 780 K) substrate temperature and 260 – 300 mtorr (300 mtorr) oxygen partial pressure. All films are about 300 nm thick. To control the oxygen content, the as-grown films were post-annealed in the furnace at the prescribed temperature and oxygen pressure condition [31]. The hole doping level of YBa$_2$Cu$_3$O$_{7-y}$ and La$_{2-x}$Sr$_x$CuO$_4$ films was carefully determined. For YBCO superconducting samples, *y* was estimated from 1- $T_c$ / $T_{c,\,max}$ = 82.6*(*y* - 0.16)$^2$, where $T_{c,\mathrm{max}}$ is the maximum transition temperature at the optimal doping level [31-33]. For heavily underdoped and undoped thin films that did not show superconductivity, *y* was determined from the thermoelectric power (TEP) at room temperature [31-33]. In addition, the *y* values estimated from these two methods (if $T_c$ was available) were cross-checked and the results were in good agreement with each other. For LSCO, the Sr concentration was quantitatively determined using Auger electron spectroscopy (AES). The actual Sr contents were consistent with the nominal $x$ values of the LSCO targets within an error of 3% or less. To compare the doping dependent RIXS spectra of YBCO and LSCO, we use the concentration of doped holes $p$ as the common parameter. For LSCO, $p$ is proposal to the amount of Sr substitution, hence $p = x$; For YBCO, $p$ is estimate according to $T_c$ [33] and the relation between $p$ and $y$ can be found in Ref. [34-36].

RIXS simulations were performed using an atomic multiplet model of $3d^9$ Cu, including $3p$ and $3d$ spin-obit coupling and the photon matrix elements defined by the experimental scattering geometry. The fraction of resonance-active $3d^9$ Cu in the YBCO chain was determined by a separate assay, which is discussed in the Supplementary Materials. This chain fraction was weighted into the spectral function of related excitations. The $3d$ crystal field was defined to match the strong tetragonal and square-pyramid orbital energy hierarchies of LSCO and YBCO, respectively (see Fig. 1(a)). Due to a different ligand environment around the YBCO chain, *e.g.* lacking an apical oxygen and with a slightly reduced Cu-O distance (from 1.91 Å to 1.86 Å) along one axis of the Cu-O plaquette plane [the symmetry breaking on $xz/yz$ orbitals is neglected here], the energetics of $dd$ excitations in the chain are distinct from the plane. Combined with the absence of ZRS in the chain, the energies of all $dd$ excitations in the simulated chain spectrum are effectively increased with respect to those in the simulated plane spectrum, see Fig. 3(c). We also note that single-magnon excitations have roughly constant energetics in the relevant doping range.

Modifications to the RIXS spectrum from hole doping were included as follows. We first introduced a near-neighbor Coulomb interaction parameter U' between the Cu $3d(x^2-y^2)$ orbital and ZRS holes on the nearest-neighbor Cu-O plane lattice sites with a mean field density $<n_{ZRS}>$. This term was added to the mean-field Hamiltonian for each bond as $H_U=U'*<n_{ZRS}>$. Secondly, we considered the transformation of *dd* excitation line shape due to the simultaneous continuum excitations of the ZRS band, representing the shake-up scattering processes induced by the creation of a localized $3d^{10}$ site in the lattice during the RIXS process [37]. The cross-section of continuum excitations ($I_c$) is proportional to the ZRS density, *e.g.* $I_c=\alpha*<n_{ZRS}>$. For simplicity, we set $\alpha =1$. These continuum excitations themselves are represented as a linearly decaying high energy tail that extends 1 eV beyond each *dd* excitation. The local spin exchange field is accounted for by coupling with an $S=2*(1-<n_{ZRS}>)$ spin moment. We used the spin exchange constants of J=105 meV and J=130 meV for the $3d(x^2-y^2)$ orbital for YBCO and LSCO, respectively [38].

### III. Results and Discussion

In Figs. 1(b) and 1(d), we show the detailed incident photon energy dependence of RIXS spectra (RIXS maps) recorded around the Cu $M_{2,3}$-resonances from undoped Mott insulator YBa$_2$Cu$_3$O$_{7-y}$ and La$_2$CuO$_4$ (LCO) films (for selected XAS spectra, see Fig. S4 in the Supplementary Materials [30]). In these RIXS maps, strong RIXS features (non-dispersive in these maps) can be seen in the 1.5 ~ 2.5 eV energy loss window when the incident photon energy is tuned to Cu $M_3$ resonance (~ 74 eV). These features are the *dd* excitations and can be identified as transitions to the unoccupied $x^2-y^2$ orbital from the occupied *xy*, *xz/yz*, and $3z^2-r^2$ orbitals, respectively (with increasing energy loss) [39]. Because of the distinct local symmetry around the central Cu site, *i.e.* pyramidal for YBCO and tetragonal for LCO, the *dd* excitations in

YBCO are located at lower energy loss and are not as well-resolved as those in LCO. Furthermore, the YBCO film emits stronger fluorescence (dispersive features in these maps), which overlaps with the *dd* excitations when incident photon energy is tuned to Cu $M_2$ resonance (~ 76.5 eV). For both samples, the localized *dd* excitations seem to display some degree of dispersion towards higher energy loss with increasing incident photon energy, suggesting that these excitations have a multi-component nature. This behavior is analogous to the phonon- and spin- dressed orbital excitations seen in other cuprates [14,35,40]. However, the expected contribution from phonons is greatly reduced at the *M*-edge [29]. This spectral profile is reproduced by our atomic multiplet calculations shown in Figs. 1(c) and 1(e), in which the effect stems from the antiferromagnetic exchange field. The energy scales for *xy*, *xz/yz*, and $3z^2$-$r^2$ orbital relative to the $x^2$-$y^2$ orbital were set as follows: for LCO, they were -1.57 eV, -1.94 eV, and -1.57 eV, respectively; for YBCO planes (chains), they were -1.49 eV (-1.65 eV), -1.76 eV (-2.16 eV), and -1.91 eV (-2.6 eV), respectively. Setting the energy scales of *dd* excitations as such yielded the best agreement between the experimental and simulated RIXS maps.

The incident photon energy and doping dependence of RIXS spectra from YBCO and LSCO are shown in Figs. 2(b) – 2(j). They are overlaid with spectra from atomic multiplet calculations (black curves). The intensity of *dd* excitations is maximal near the Cu $M_3$ resonance around 74 eV. The agreement between the experimental (colored lines) and simulated (black lines) RIXS spectra for low doping films also becomes better around the resonance (green curves, panels (b) and (f)). When the hole doping level is increased, these *dd* excitations become broader; in the meantime, an extended spectral tail becomes visible on the higher energy loss side (panels (d) and (i)). Further increasing the hole doping level to reach the superconducting regime (panels (d), (e), (h), (i), and (j)) leads to the smeared *dd* excitations that are too broad to be resolved in the spectra. This asymmetrical broadening effect is approximated by the ZRS continuum excitation tails within the calculations and is far more noticeable in LSCO than YBCO. This is possibly related to their contrasting hole doping mechanism ($CuO_2$ chain as the hole reservoir in YBCO versus the cation substitution in LSCO). Despite the complexity introduced by high energy continuum excitations, the low energy onset of *dd* excitations remains sharply defined and is expected to represent the energy needed to create an isolated $3d_{xy}$ *dd* excitation with no simultaneous continuum excitation [37].

In Fig. 3(a), we show the RIXS spectra taken at 74.5 eV incident photon energy from different YBCO films, with the intensity at 1.6 eV energy loss rescaled to 1 for the ease of comparison. Upon hole doping, the tail structure at higher energy loss side remains nearly unchanged whereas the leading edge at lower energy loss side shifts monotonically towards the elastic peak (see blue open circles in Fig. 3(e)). Unlike LSCO, there are two different Cu sites in YBCO: $CuO_2$ planes and $CuO_2$ chains with Cu in primarily $3d^9$ and $3d^{10}$ configurations, respectively. One might speculate that this leading edge shift is related to the hole doping into $CuO_2$ chains that opens up available transitions; however, our calculations show that the chain-related features appear at higher energy loss around 1.775 eV and 2.225 eV in the RIXS spectra, and should be weakly visible due to the low density of RIXS-active $3d^9$ sites in the chain structure (see Experiments section and Supplementary Materials [30]). Moreover, a very similar leading edge shift phenomenon can also be found in LSCO films, whose RIXS spectra taken at the same photon energy are shown in Fig. 3(b). For LSCO, one can see that hole doping leads to the significant broadening of *dd* excitations and the extension of higher energy loss tail, indicating a noticeable damping of these localized excitations by the coupling to other electronic degrees of freedom; however, the leading edge shows the consistent shifting trend as the YBCO films (red open circles in Fig. 3(e)). One may speculate that this effect can be explained by broadening the *dd* excitations, but this scenario can be ruled out (see Fig. S5 in Supplementary Materials [30]). In that regard, the observed leading edge shift is intrinsic and related to the hole doping to the $CuO_2$ planes.

To obtain insight into the origin of this phenomenon, we have carried out atomic multiplet calculations with incorporation of Coulomb attraction between the Cu $3d(x^2-y^2)$ electron and the doped holes in the surrounding oxygens [15]. Note that with hole doping, the local crystal field around the Cu site in $CuO_2$ plane can change and lead to the energy shift in the leading edge of *dd* excitations. However, the energy scale associated with the crystal field is often fairly large, on the order of 100 meV, and is usually accompanied by structural response in terms of changing the crystal symmetry and/or lattice parameters. Such a structural response is not a common feature between these two cuprate families; therefore, we focus on this Coulomb interaction in a mean-field picture.

We look at how the energy of Cu $3d(x^2-y^2)$ orbital is influenced by the dopant-induced changes to the charge density distribution. The presence of ZRS on four nearest neighbors of the scattering site reduces the Coulomb and spin superexchange energy of the vacant $3d(x^2-y^2)$ state. This changes our mean field Hamiltonian from $E_{x2-y2}$ to $E = E_{x2-y2} - <n_{ZRS}>*4(U'+J/2)$. This doping dependence in the $3d(x^2-y^2)$ state directly modifies the energy of *dd* excitations, causing their leading edge to shift towards smaller energy loss (see Fig. 3(d)). The existence of vacancies in the ZRS band also enables gapless intra-band continuum excitations that occur simultaneously with *dd* excitations, resulting in a higher energy tail next to these *dd* excitations. To avoid this complexity, our investigation focuses only on the leading edge of the lowest energy *dd* excitation with *xy* symmetry, which does not overlap with other *dd* excitations or with the signal attributed to CuO$_2$ chains in YBCO (see Fig. 3(c)). In Fig. 3(e), we plot curves with several U' values from 0.1 eV to 0.2 eV and one can see that 0.15 eV gives the best agreement with the experimental data.

Before proceeding, we note that the ZRS originating on the scattering site is assumed not to contribute to the leading edge of *dd* excitations because (1) the single-plaquette RIXS process typically does not leave the ZRS mode in the low energy singlet sector; (2) in a larger many-body picture, such a ZRS is significantly perturbed in both the intermediate and final state configurations of the RIXS process; and (3) the singlet binding energy of a ZRS mode is significantly reduced by the presence of a *dd* excitation, and this will detune excitations with on-site ZRS modes away from the specific features that we are tracking. In addition, excitations in which a core electron directly enters the ZRS band are not considered because they will result in a higher energy $3d^8$ configuration that is outside the measured energy loss window.

We now discuss the meaning of these findings. The energy scales of *dd* excitations have significant influence on the material dependence of $T_c$ [12,13]. One proposed connection is the energy of $3z^2$-$r^2$ orbital relative to the $x^2$-$y^2$ orbital because a smaller energy difference can lead to a larger component in the $3z^2$-$r^2$ orbital (and the hybridized O $2p_z$ orbital), weakening the *d*-wave superconductivity. Due to the simplicity and higher energy resolution of Cu *M*-edge RIXS, we are able to determine the energy of $3z^2$-$r^2$ orbital $E_{3z2\text{-}r2}$ (relative to the energy of $x^2$-$y^2$ orbital) for LCO and YBCO. Indeed, $E_{3z2\text{-}r2}$ (= 1.57 eV) of LCO is significantly smaller than that of the YBCO (1.91 eV), consistent with their lower $T_c$. In the literature, $E_{3z2\text{-}r2}$ of LCO is probably more investigated both experimentally and theoretically than that of YBCO. For example, Ref. [37] suggested a value of $E_{3z2\text{-}r2}$ = 1.70 eV for LCO, and the magnitudes of $E_{3z2\text{-}r2}$ and $E_{xy}$ are comparable. Both statements are consistent with the present findings for LCO. It is also noted that the observed $E_{3z^2-r^2}$ for LCO is larger than the calculated $E_{3z^2-r^2} = 0.91$ eV in Ref. [13]. However, the qualitative trend of $T_c$ vs. $E_{3z^2-r^2}$ in Ref. [37] is in accord with our results.

We have also acquired an estimate of the interaction strength between $3d(x^2-y^2)$ electrons and doped holes on the surrounding copper oxygen plaquette with unprecedented accuracy. This "direct" observation would be impossible without the energy resolution available at $M$-edge RIXS. As mentioned above, the correct value of U' has a profound impact on both the fundamentals of Hubbard physics and the interpretations of optical spectroscopy data in undoped and doped Mott insulators. For example, the electron-hole Coulomb attraction is at the heart of charge density wave (CDW) phenomena, which is a generic feature among cuprate families. The ability to determine the U' value (hence $U_{pd}$, see below) may help understand the two-dimensional (2D) CDW in YBCO, but the observation of a 3D CDW in YBCO film will likely call for a more extended model such as a larger multiband Hubbard model that incorporates the apical orbitals to transmit the charge order correlation through the chain layers [41].

Comparing the present value of U' = 0.15 eV with earlier literature requires sophisticated treatment as the interaction is commonly parameterized in terms of $U_{pd}$, the Coulomb interaction term between a Cu $3d(x^2-y^2)$ electron and a sigma-bound O $2p$ orbital. The highest energy ZRS modes are thought to be shared roughly equally between Cu $3d$ and O $2p_\sigma$ orbitals, and the single-site representation of a ZRS mode is split between 4 oxygen atoms. [42] As such, only ~1/8 of the ZRS hole density is expected to occupy the neighboring $2p_\sigma$ orbitals, and the nominal value of $U_{pd}$ is roughly 8 times larger than our quantity U', or $U_{pd}$~1.2 eV. This value is agreement with the numerical estimate of 1.3 eV$\geq U_{pd}>$0.6 eV in Ref. [43], and with rough parameterizations of $U_{pd} \geq 0.5$ eV in other work [17,18,20].

Framing U' as the electron-ZRS interaction parameter has the advantage of being more directly relevant to the single band Hubbard model. It also highlights that U' is an effective parameter that cannot necessarily be evaluated very accurately from electrostatics alone. However, it is important to note several approximations used in our calculations may influence this estimate. Firstly, our description of local exchange field as coupling to an $S=2*(1-<n_{ZRS}>)$ neighboring spin moment is a very rough approximation. If this spin effect were neglected altogether, the U' value would increase by 9% to 0.16 eV. Secondly, we have ignored the effect of neighboring ZRS modes on the $3d(xy)$ orbital energy. This omission causes our U' value to be an underestimate, as it is in fact the difference between the ZRS Coulomb interaction constants for the $x^2$-$y^2$ orbital and the $xy$ orbital ($U'=U'_{x2-y2}-U'_{xy}$). Lastly, we have approximated that the excitations at the leading edge of $3d_{xy}$ dd excitation can only occur when there is no on-site ZRS mode. This approximation is well motivated, but will nonetheless lead to some undercounting of the local hole density. This in turns causes an overestimation of U'. The second and third effects act on U' with opposite sign, and are both likely to contribute to an error at the ~10% level. Intriguingly, the observed leading edge shift of the excitations due to hole doping in LSCO and $Sr_{14-x}Ca_xCu_{24}O_{41}$ coincides with that in the present RIXS case [19,43].

## IV. Summary

In summary, we have performed RIXS measurements on cuprate films in the extreme ultraviolet energy regime. From the energy dependence of RIXS map, we are able to precisely determine the energetics of $3d$ orbitals (with respect to $x^2$-$y^2$ orbital) in YBCO and LCO Mott insulators. Our results support that the energy of $3d$ $z^2$-$r^2$ orbital has a significant impact on $d$-wave superconductivity. The linear shift of leading edge of $dd$ excitations towards smaller energy loss as a function of hole doping is observed in both systems. Such phenomenon can be reproduced by introducing an attractive Coulomb interaction between Cu $3d(x^2-y^2)$ electron and doped holes on the surrounding oxygens in atomic multiplet calculations. This attractive Coulomb interaction can link to the parameter $U_{pd}$, which is an important factor in many theoretical models, but has not been directly measured before.

The Advanced Light Source is supported by the Director, Office of Science, Office of Basic Energy Sciences, of the U.S. Department of Energy under Contract No. DE-AC02-05CH11231. This work was supported by National Science and Technology Council (NSTC) of Taiwan under Grants No. 108-2112-M-009-009, No. 110-2634-F-009-026, and No. 107-2199-M-009-002. The work of J.Y.L was also financially supported by center for Emergent Functional Matter Science of National Yang Ming Chiao Tung University from The Featured Areas Research Center Program within the framework of the Higher Education Sprout Project by Taiwan Ministry of Education. We would like to thank C.-C. Chen for fruitful discussions. Research at New York University was supported by the National Science Foundation under Grant No. DMR-2105081

E-mail: shih.huang@psi.ch ; lawray@nyu.edu;  ago@nycu.edu.tw

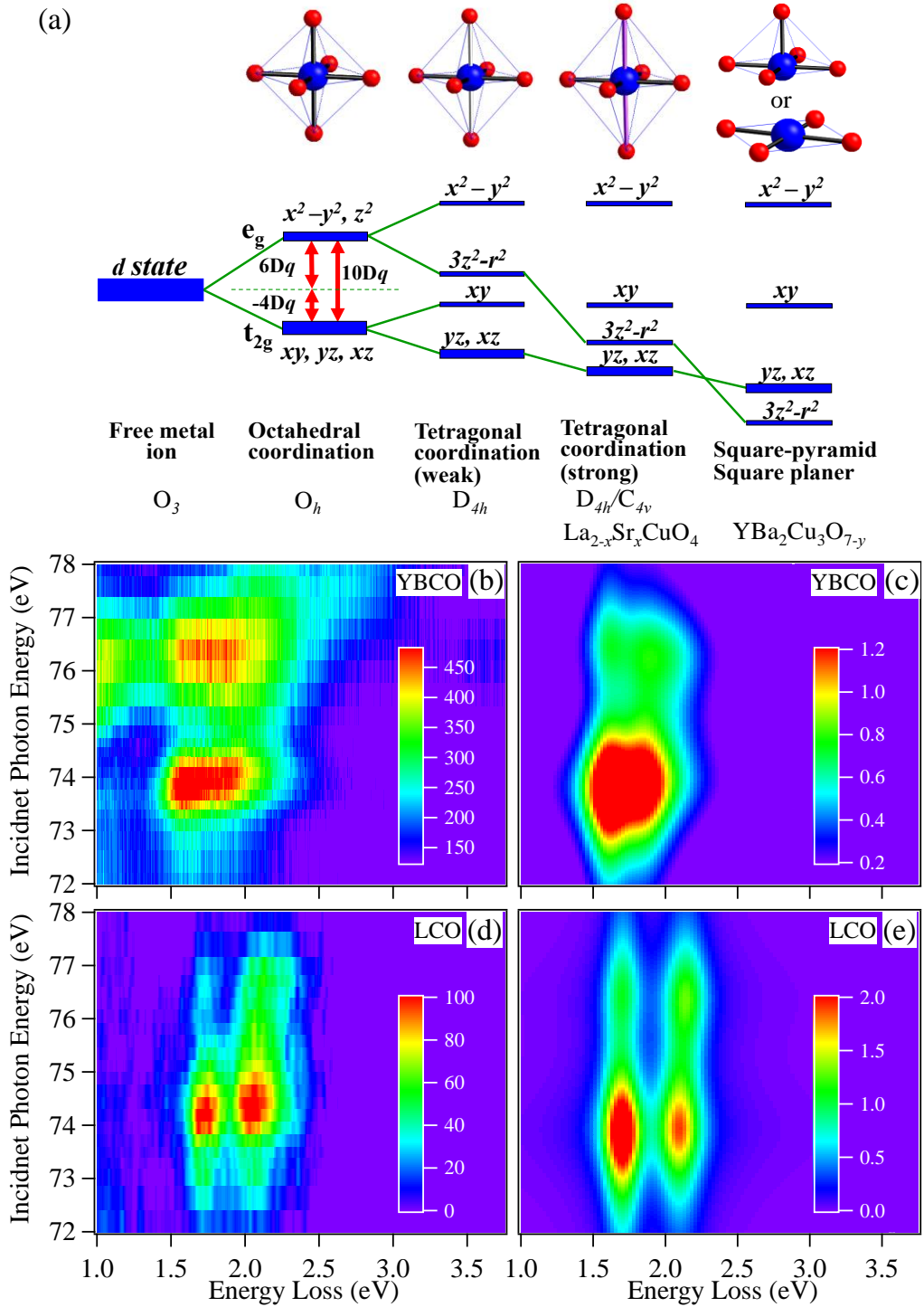

FIG 1. (a) Crystal field splitting from octahedral to tetragonal, square pyramidal, and square planar symmetry. For atomic multiplet calculations, we used the strong tetragonal and square pyramidal symmetry for $La_{2-x}Sr_xCuO_4$ and $YBa_2Cu_3O_{7-y}$, respectively. (b, d) The experimental RIXS maps plotted in energy loss scale for (b) $YBa_2Cu_3O_6$ ($p=0$, Mott insulator), and (d) $La_2CuO_4$ around Cu $M_{2,3}$-edge resonances. (c, e) The simulated RIXS maps for (c) $YBa_2Cu_3O_6$ Mott insulator and (e) $La_2CuO_4$.

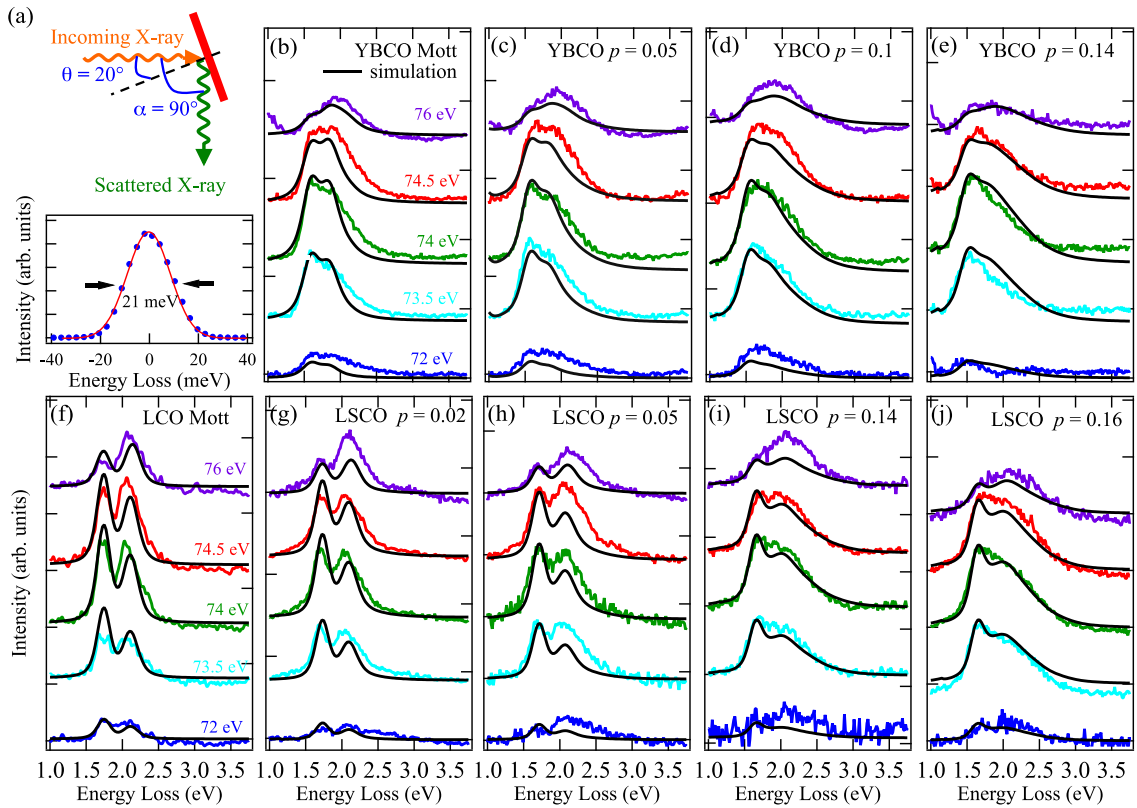

FIG 2. (a) Schematic illustration of experimental geometry. The energy resolution determined from the full width at half maximum of elastic peak is 21meV (blue dots: data; red curve: Gaussian fit). Incident photon energy dependent RIXS spectra of $YBa_2Cu_3O_{7-y}$ with $p =$ (b) 0, (c) 0.05 (no $T_c$), (d) 0.1 ($T_c = 50$ K), (e) 0.14 ($T_c = 90$ K) and $La_{2-x}Sr_xCuO_4$ with $p =$ (f) 0, (g) 0.02 (no $T_c$), (h) 0.05 ($T_c = 5$ K), (i) 0.14 ($T_c = 35$ K), and (i) 0.16 ($T_c = 40.1$ K). In panels (b) – (j), the blue, cyan, green, red, and purple curves are spectra taken at incident photon energy of 72, 73.5, 74, 74.5, and 76 eV, respectively. Black curves: simulation.

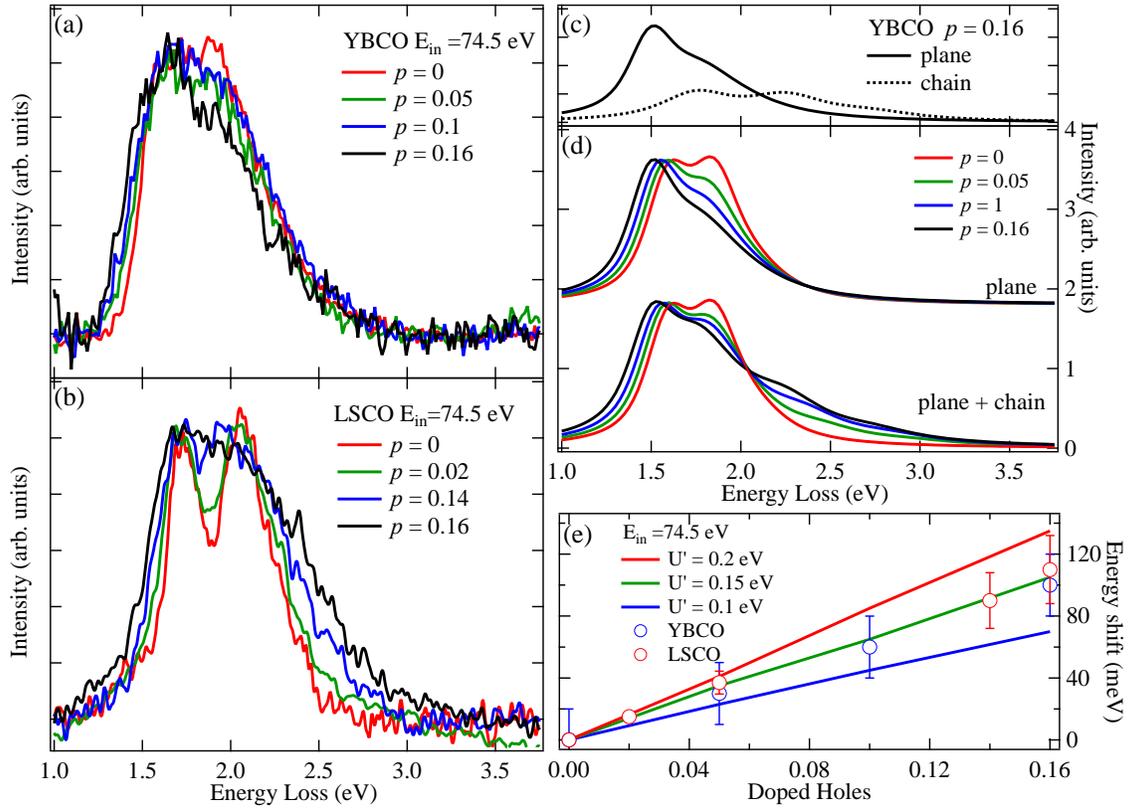

FIG 3. Doping dependent RIXS spectra of (a) YBa$_2$Cu$_3$O$_{7-y}$ and (b) La$_{2-x}$Sr$_x$CuO$_4$ recorded at 74.5 eV incident photon energy. The spectra are normalized at 1.6 eV energy loss. (c) Simulated RIXS spectrum of YBCO showing the chain (dashed line) and plane (solid line) contributions to the *dd* excitations. (d) Simulated RIXS spectra for YBCO, the energy of *x²-y²* orbital of different doping is modified with $E_{x2-y2} - \langle n_{ZRS}\rangle \ast 4(U'+J/2)$, showing the monotonic shift of the leading edge of *dd* excitations. The doped holes contribute to $\langle n_{ZRS}\rangle$. Note that with increasing the hole doping, the spectral weight of high energy tail also increases due to doped holes in CuO$_2$ chain. (e) Overlay of the leading edge energy shift from experimental (red open circles: YBCO; blue

open circles: LSCO) and simulated (lines) RIXS spectra with different Coulomb attraction values U'.

Supplementary Materials for

**Precise *dd* excitations and commensurate intersite Coulomb interactions in the dissimilar cuprate YBa$_2$Cu$_3$O$_{7-y}$ and La$_{2-x}$Sr$_x$CuO$_4$**

**Resonant inelastic x-ray scattering (RIXS) measurements**

To reveal the subtle change in the energy loss of dd excitations, a precise determination of energy of these excitations with respect to the strong elastic peak is crucial. We looked at elastic peak position on the detectro as we vary the photon energy and used a 3rd order polynomial fitting to obtain a conversion between pixel number and photon energy, thereby converting the horizontal axis of spectra to energy loss. This approach allows us to determine the energy loss scale better than 2.5 meV roughly 10% of the energy resolution as determined from the full width of half max (FWHM) of the elasctic peak.

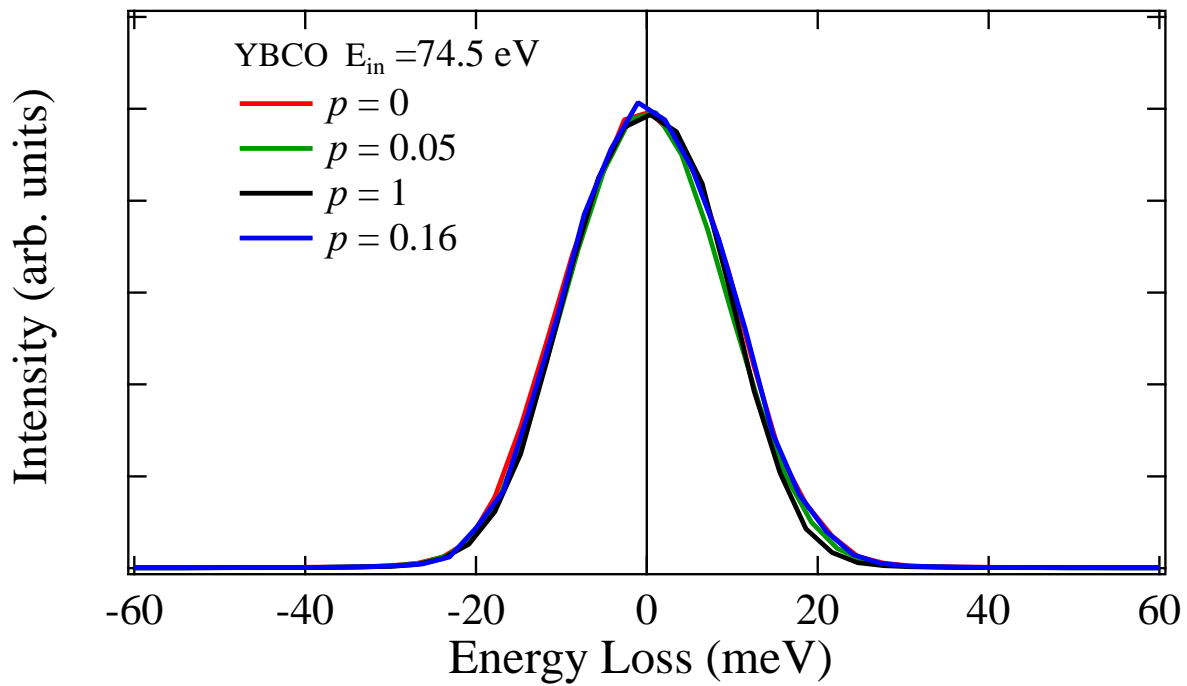

**Figure S1.** The normalized elastic peak of $YBa_2Cu_3O_{7-y}$ with different doping. The excitation photon energy is 74.5 eV.

## The evolution of $Cu^{1+}$ fraction in $YBa_2Cu_3O_y$

In the atomic multiplet calculations, we utilize the information of the fraction of $3d^9$ $Cu^{1+}$ in the YBCO chain. $Cu^{1+}$ manifests itself in the Cu $L_{III}$-edge XAS as the feature at 934 eV shown in Fig. S2(a). The Cu $L_{III}$ edge is fitted with four Gaussian peaks respectively assigned to $Cu^{2+}$, the ligand, $Cu^{1+}$, and a high energy feature of the unknown origin. Fig. S2 shows the examples of $p = 0$ and 0.08 from batch A. The $p = 0$ sample corresponds to $YBa_2Cu_3O_6$, the Mott insulator mother compound, which is presumed to have 100% $Cu^{1+}$ in the chain structure. Assuming the number of $Cu^{1+}$ in the chain is proportional to the spectral weight of the $Cu^{1+}$ peak in Fig. S2 for $YBa_2Cu_3O_{7-y}$, the normalization of the spectral weight $S_{Cu^{1+}}(p)/S_{Cu^{1+}}(p=0)$ leads to the evolution of $Cu^{1+}$ fraction as shown in Fig. S3. We perform this practice with two independent batches of $YBa_2Cu_3O_{7-y}$ and the results are consistent.

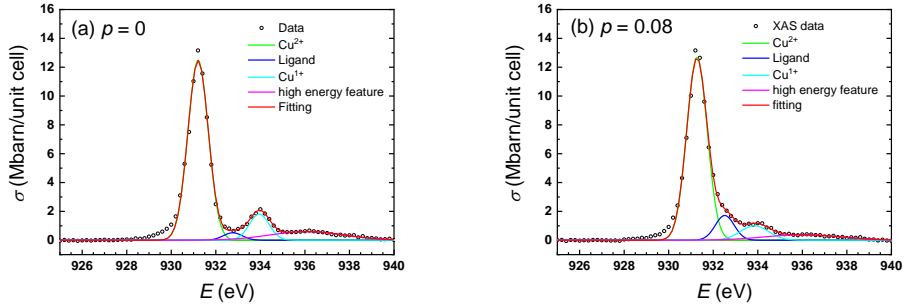

Fig S2 Cu $L_{III}$-edge XAS of $YBa_2Cu_3O_{7-y}$. (a) $p = 0$; (b) $p = 0.08$.

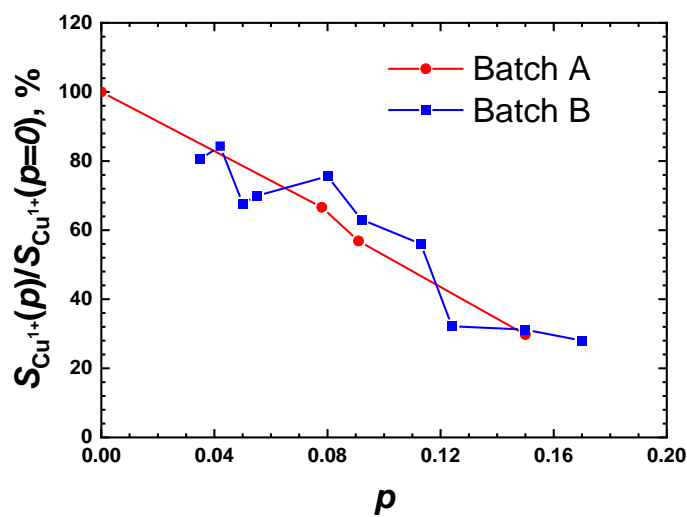

Fig. S3 The evolution of $Cu^{1+}$ fraction with the doping level $p$ in $YBa_2Cu_3O_{7-y}$.

**Selected X-ray abosprtion spectra of $YBa_2Cu_3O_{7-y}$ with different doping level and $La_2CuO_4$**

In Figure S4, we show selected XAS spectra (recorded in total electron yield mode) of $YBa_2Cu_3O_{7-y}$ (YBCO) with different hole doping level and $La_2CuO_4$ at room temperature. As one can see, despite with different hole doping levels, the spectral lineshape remains nearly the same. Unlike the *L*-edge XAS of this compound, there is no identifiable feature in the *M*-edge XAS spectrum that corresponds to the absorption of $CuO_2$ chain [33]. In that regard, it is unlikely that one can use the XAS spectrum in the atomic multiplet calculations to distinguish the chain and plane contributions.

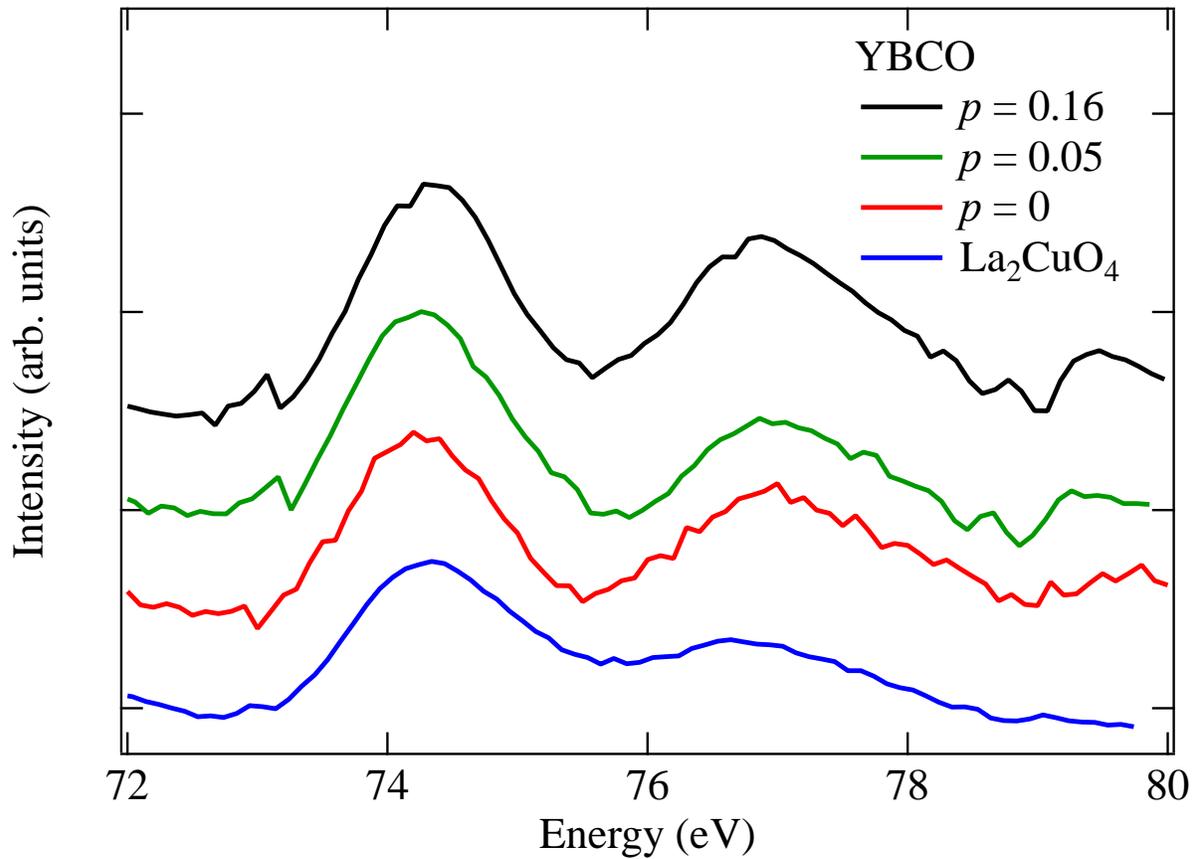

Figure S4: Selected XAS spectra of YBCO recorded in total electron yield mode at room temperature. Here, $p$ denotes the hold doping level.

**Braodening effect on RIXS spectra**

In Figure S5, we show the broadening of the undoped (Mott) LSCO spectrum (solid green line) by convolving it with a Gaussian function. Compared to the x=0.16 spectrum (solid blue line), here are characteristic differences:

- The intensity of broadened spectra around 2 eV energy loss is consistently larger than the $x = 0.16$ spectrum, irrespective to the broadening width (20 to 200 meV).
- The intensity of broadened spectra around 1.6 eV energy loss is consistently smaller than the $x = 0.16$ spectrum, and
- With increased broadening width, although the high energy loss tail progressively approaches that in $x = 0.16$ spectrum, the slope of leading edge becomes smaller.

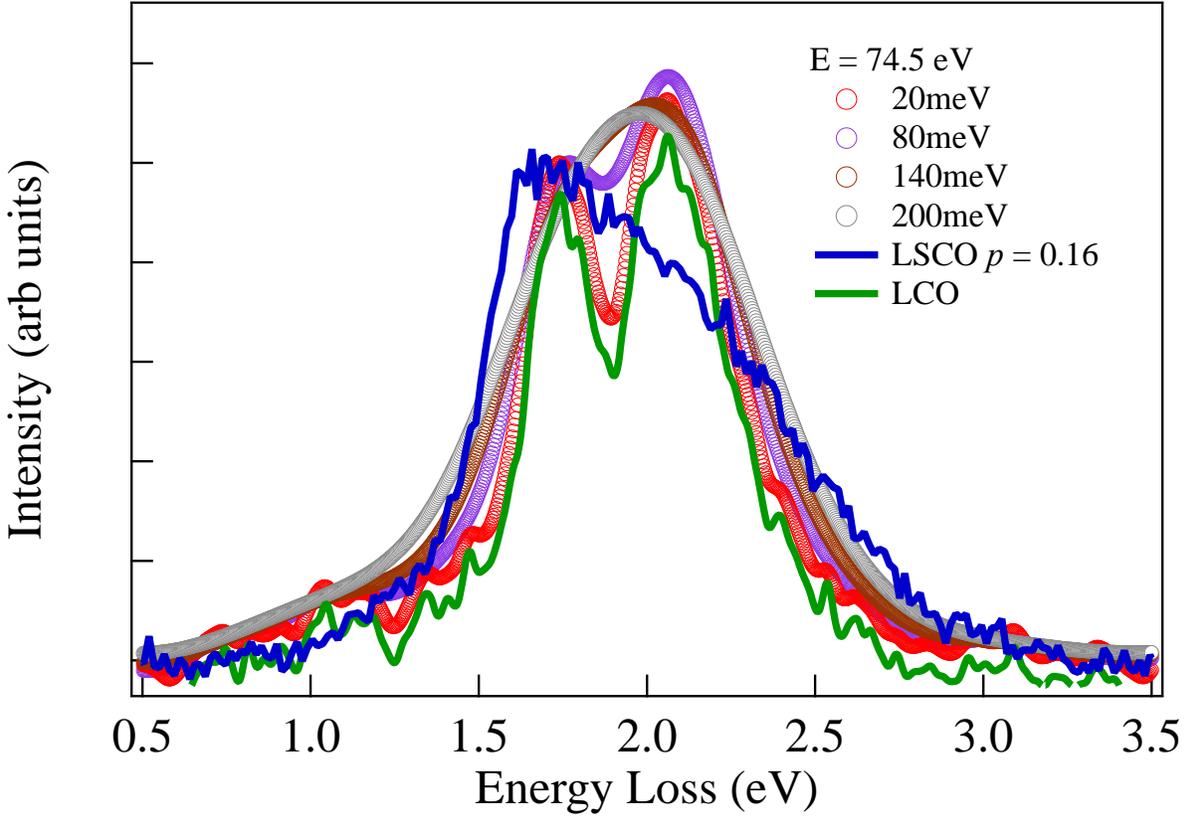

Figure S5: The LCO spectrum with different Gaussian broadening (20meV, 80meV, 150meV and 200meV).

The leading edge of *dd* excitations is generically not very susceptible to broadening from doped charge carriers because such a broadening mostly comes in a form of higher energy (trailing edge) continuum excitation due to enhanced interactions [37].